\documentclass[12pt]{article}

\usepackage{amsmath}
\usepackage{amssymb}
\usepackage{graphicx}
\usepackage{subfigure}
\usepackage{epsfig}
\usepackage{alltt}
\usepackage{boxedminipage}
\usepackage{float}

\newenvironment{code}{
\footnotesize
\begin{alltt}
\begin{center}
\begin{boxedminipage}{1.0\textwidth}
}{
\end{boxedminipage}
\end{center}
\end{alltt}
}

\newcommand{\mbs}[1]{\bf{#1}}

 \def\bt{{\mbs{t}}}

\def\ttet{{{\tt tetrahedron}}}
\def\ttets{{{\tt tetrahedra}}}
\def\tfac{{{\tt facet}}}
\def\tfacs{{{\tt facets}}}
\def\tseg{{{\tt segment}}}
\def\tsegs{{{\tt segments}}}

\title{\bf {An Efficient Adaptive Procedure
for Three-Dimensional Fragmentation Simulations}}

\date{}

\author{
{\bf {Anna Pandolfi}}\\
{\small Dipartimento di Ingegneria Strutturale}\\
{\footnotesize Politecnico di Milano, 20133 Milano, Italy}\\
{\footnotesize e-mail: pandolfi@stru.polimi.it}\\
\\
{\bf {Michael Ortiz}}\\
{\footnotesize Graduate Aeronautical Laboratories}\\
{\footnotesize California Institute of Technology, Pasadena,
CA 91125, USA}\\
{\footnotesize e-mail: ortiz@aero.caltech.edu}
{\small Adaptive remeshing, Cohesive elements, 3D Finite elements.} \\
}

\begin{document}

\maketitle

\subsection*{Abstract}
\label{abstract}

{\small We present a simple set of data structures, and a
collection of methods for constructing and updating the
structures, designed to support the use of cohesive elements in
simulations of fracture and fragmentation. Initially all interior
faces in the triangulation are perfectly coherent, i.~e.,
conforming in the usual finite element sense. Cohesive elements
are inserted adaptively at interior faces when the effective
traction acting on those face reaches the cohesive strength of
the material. The insertion of cohesive elements changes the
geometry of the boundary and, frequently, the topology of the
model as well. The data structures and methods presented here are
straightforward to implement and enable the efficient tracking of
complex fracture and fragmentation processes. The efficiency and
versatility of the approach is demonstrated with the aid of two
examples of application to dynamic fracture.}

\section{Introduction}
\label{Introduction}

The essential interplay between geometry and mechanics comes into
sharp focus in applications where the topology of the domain may
change, oftentimes extensively, in the course of calculations.
Fragmentation, which may result in a runaway proliferation of
bodies in the form of fragments \cite{field:1989, kipp:1993,
woodward:1994, piekutowski:1995}, provides a case in point.

Camacho and Ortiz \cite{camacho:1996, ortiz:1996} in two          
dimensions, and Pandolfi {\it et al.} \cite{pandolfi:1999,   
ruiz:2000, ruiz:2001} in three                                    
dimensions, have established the feasibility of: i) accounting
explicitly for individual cracks as they nucleate, propagate,
branch and possibly link up to form fragments; and ii) simulating
explicitly the granular flow which ensues following widespread
fragmentation.  In this approach, cracks are allowed to form and
propagate along element boundaries in accordance with a
cohesive-law model \cite{ortiz:1993, xux:1994}. Clearly, it is    
incumbent upon the mesh to provide a rich enough set of possible
fracture paths, an issue which may be addressed within the
framework of adaptive meshing.  In contrast to other approaches
\cite{ortiz:1993, xux:1994} which require interfacial elements    
to be inserted at the outset along potential fracture paths,
Camacho and Ortiz \cite{camacho:1996}, and Pandolfi {\it et al.}
\cite{pandolfi:1999, ruiz:2000, ruiz:2001},                       
adaptively create new surface as required by the
cohesive model by duplicating nodes along previously coherent
element boundaries and inserting surface-like \emph{cohesive
elements} which encapsulate the fracture behavior of the solid.
These elements are surface-like and are compatible with general
bulk finite element discretizations of the solid, including those
which account for plasticity and large deformations.

Pandolfi and Ortiz \cite{pandolfi:1998} have given an enumeration of the
ways in which the topology and geometry of a three-dimensional
finite-element model may evolve as a consequence of fracture and
fragmentation, and have described the actions which may be taken
in order to update the boundary representation, or \emph{Brep}, of
the solid. Maintaining an up-to-date Brep is of the essence when
meshing methods such as the advancing front \cite{peraire:1987,
peraire:1988, lohner:1988} are utilized \cite{radovitzky:2000}. The Brep may 
also assist in the implementation of contact algorithms \cite{kane:1999}.   
The geometrical framework proposed by Pandolfi and Ortiz \cite{pandolfi:1998}
has extensively utilized in a broad range of applications
involving fracture and fragmentation \cite{pandolfi:1999, ruiz:2000,        
ruiz:2001}.

The Brep of a solid can become inordinately complex, and thus
computationally costly to maintain, in applications involving
profuse fragmentation. In addition, some meshing algorithms,
e.~g., those based on subdivision \cite{molinari:2001}, and contact         
algorithms (e.~g., \cite{pandolfi:2001}) do not require a full              
Brep for their implementation. In these cases,
much simpler data structures suffice in order to account for
fragmentation processes.

The purpose of this article is to present one such set of data
structures, and a suite of methods for constructing and updating
the structures. The data structures and methods are
straightforward to implement and enable the efficient tracking of
complex fracture and fragmentation processes. We also present two
examples of application to dynamic fracture which illustrate the
uncanny ability of the method to represent intricate geometrical
and topological transitions resulting from crack branching, the
nucleation of surfaces and interior cracks, crack coalescence,
the detachment of fragments, and other effects.

The organization of the paper is as follows. In Section
\ref{datastructures} we introduce three simple data structures
where all necessary information pertaining to tetrahedra, faces
and edges is stored. In Section \ref{fragmentation} we present
the suite of methods for evolving the data structures in response
to fragmentation. Finally, in Section \ref{examples} we present
examples of application which demonstrate the scope and the
versatility of the approach.

\section{Topological Data Structures for a 3D finite element model}
\label{datastructures}

\begin{figure}[htp]
\begin{center}
\mbox{
\subfigure[]{\epsfig{figure=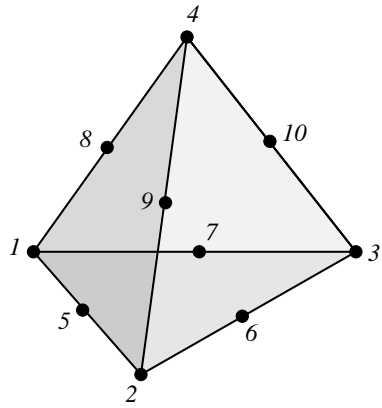,height=2.2in}}
} \mbox{
\subfigure[]{\epsfig{figure=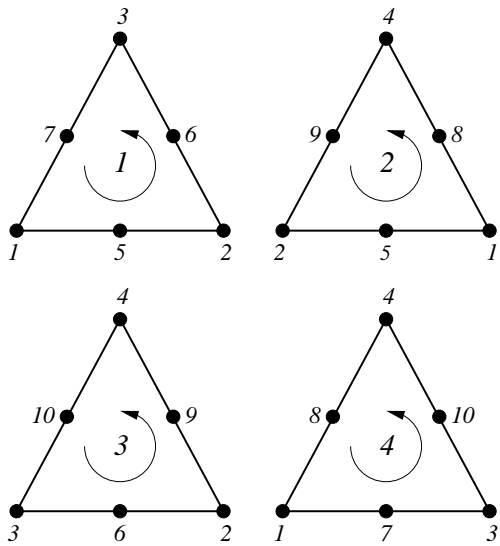,height=2.8in}}
\hspace {0.8in}
\subfigure[]{\epsfig{figure=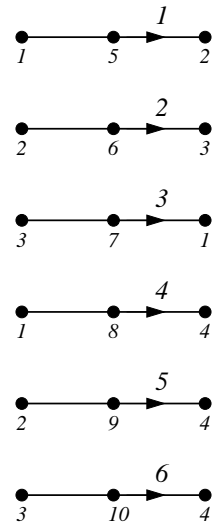,height=2.8in}}
} \vspace{0.1in} \caption[]{\small (a) A 10-nodes \ttet; (b)
description of its four \tfacs; (c) description of its six
segments.} \label{tetraelements}
\end{center}
\end{figure}

We shall restrict our attention to tetrahedral triangulations and
regard the computational model as a three-dimension simplicial
complex \cite{requicha:1980, mantyla:1988, hoffmann:1989}.        
In addition, we focus on the case of quadratic
interpolation and, hence, every tetrahedron in the triangulation
gives rise to a ten-node tetrahedral finite-element (e.~g.,
\cite{radovitzky:2000}). Extensions to higher-order elements are  
straightforward but will not be pursued here. The local nodal
numbering convention for the tetrahedral elements adopted here is
shown in Fig.~\ref{tetraelements}a. Each tetrahedron is bounded
by exactly four triangular faces, Fig.~\ref{tetraelements}b, and
six edges, Fig.~\ref{tetraelements}c. The faces and the edges can
be oriented consistently by an appropriate ordering of the nodes,
Figs.~\ref{tetraelements}b and Fig.~\ref{tetraelements}c. A face
is bounded by exactly three edges. Each face is incident on
exactly one tetrahedron, if the face is on the boundary of the
body, or two tetrahedra, if the face is in the interior. Edges
are incident on rings of varying numbers of tetrahedra. Likewise,
a variable number of faces may be adjacent to an edge.

\begin{figure}[H]
\begin{center}
\subfigure[]{\epsfig{figure=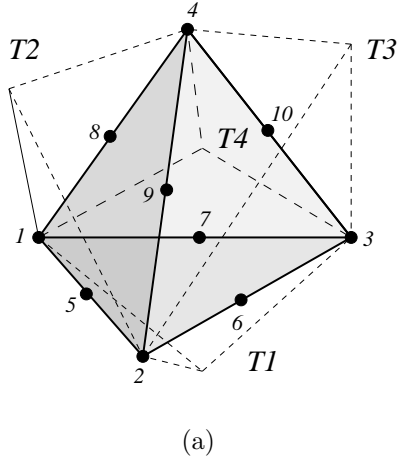,height=2in}}
\vspace{0.1in} \caption[]{\small Tetrahedron connected to four
adjacent tetrahedra.} \label{tetrainmesh}
\end{center}
\end{figure}

We begin by introducing three data structures which collect these
data and relationships, namely, the \ttets, \tfacs~ and \tsegs~
structures. In describing these structures, we adopt C syntax for
definiteness. The data stored in the \ttets~ structure consists of
the number of the element, the element connectivity, pointers to
its six \tsegs, pointers to its four \tfacs, and pointers to the
four adjacent \ttets, Fig.~\ref{codetetra}. Some of these
pointers can be null. For instance, if a tetrahedron is incident
on the boundary one or more of the pointers to adjacent
tetrahedra are null. With a view to facilitating fragmentation
simulations based on the use of cohesive elements \cite{pandolfi:1998}, we 
include in the \ttets~ structure the number of the adjacent
cohesive elements and the side of the cohesive elements incident
on the tetrahedron.

\begin{figure}[H]
\begin{center}
\begin{code}

/*--------------------------------------------------------------*
 * Tetrahedron data structure                                   *
 *--------------------------------------------------------------*/
typedef struct tetra
\{
  struct tetra   *Link, *Rink;  /*{\it Links }*/
  int             el;           /*{\it Element no. }*/
  int             N[10];        /*{\it Nodes }*/
  struct segment *S[6];         /*{\it Pointers to Segments }*/
  struct facet   *F[4];         /*{\it Pointers to Facets }*/
  struct tetra   *T[4];         /*{\it Pointers to Tets }*/
  int             C[4];         /*{\it Cohesive element }*/
  int             L[4];         /*{\it Coh. el. side: 0=bottom, 1=top }*/
\} Tetra;

\end{code}
\vspace{0.1in}
\caption[]{\small Description of the \ttet~ data structure.}
\label{codetetra}
\end{center}
\end{figure}

\begin{figure}
\begin{center}
\mbox{
\subfigure[]{\epsfig{figure=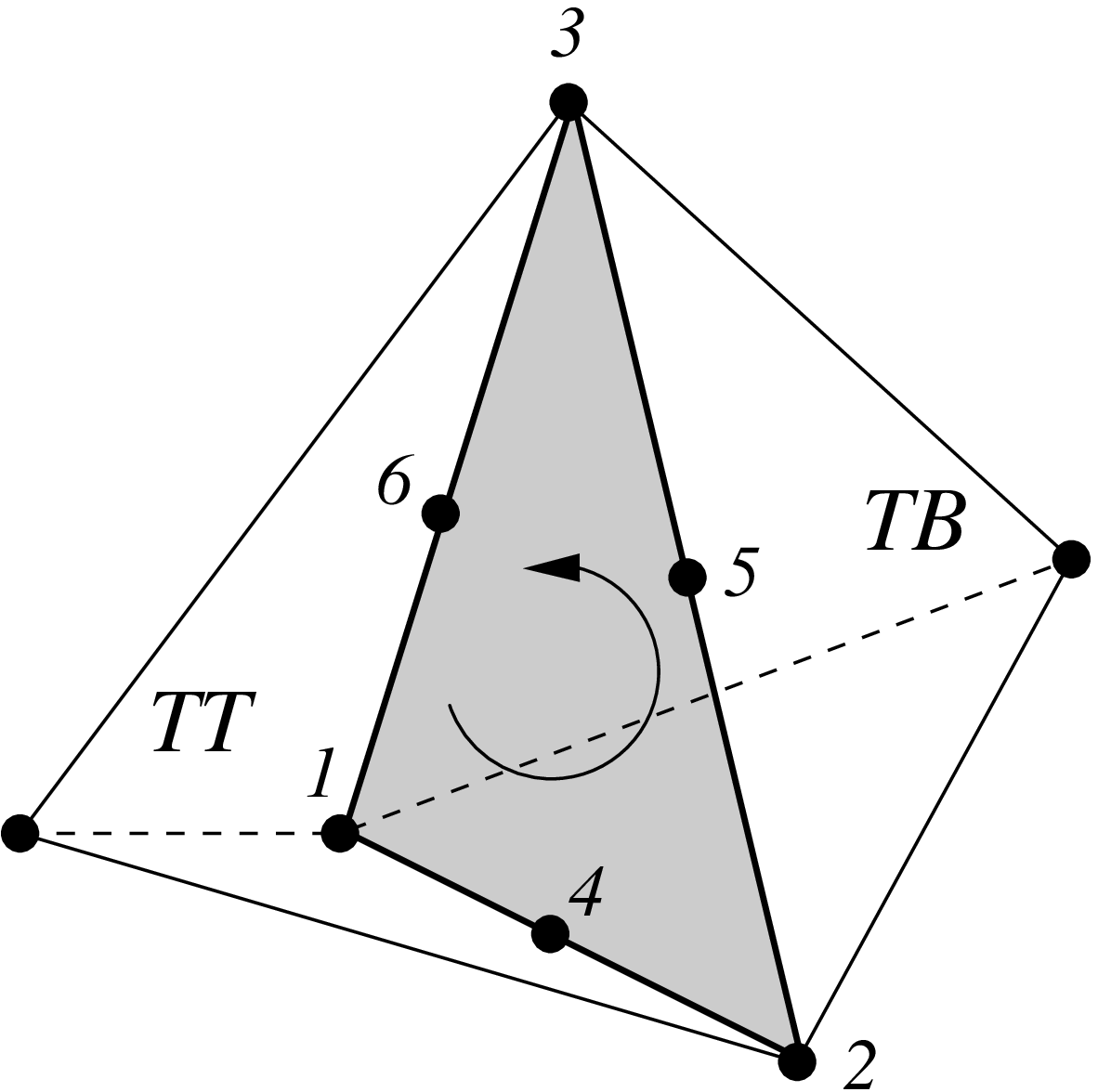,height=1.8in}}
\hspace {0.8in}
\subfigure[]{\epsfig{figure=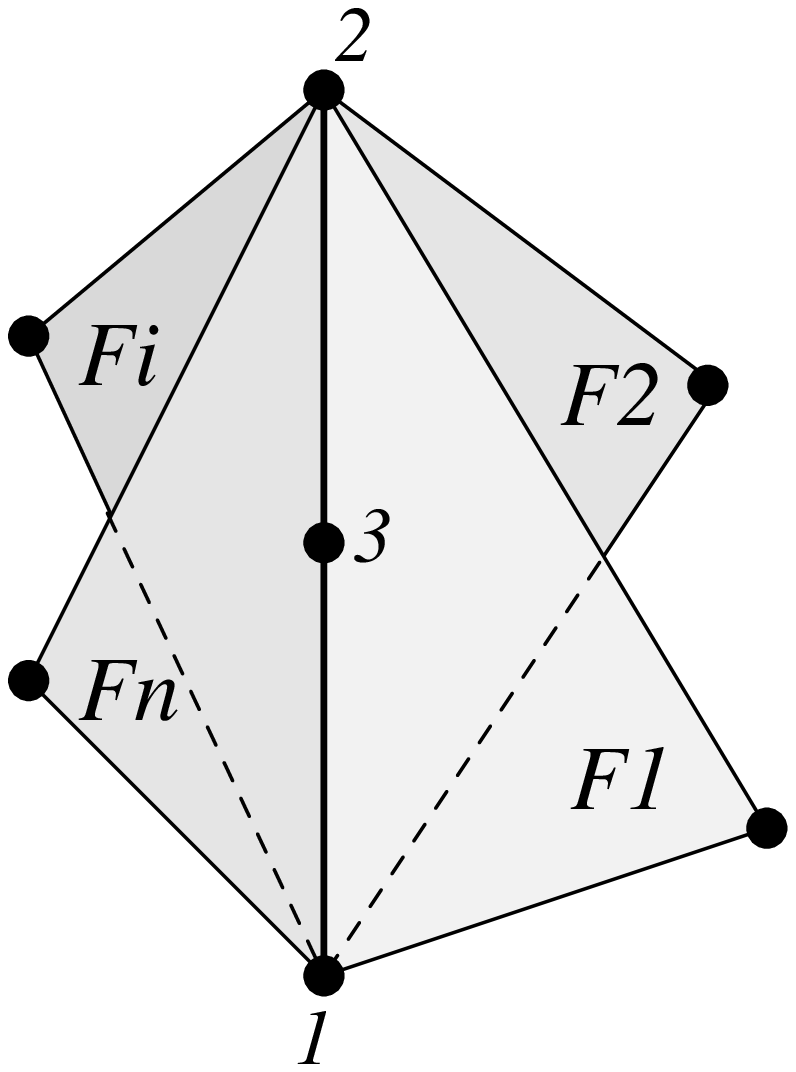,height=1.8in}}
}
\vspace{0.1in}
\caption[]{\small
(a) A \tfac~ connected to two \ttets;
(b) a \tseg~ connected to several \tfacs.}
\label{assembledelements}
\end{center}
\end{figure}

The data stored in the \tfac~ structure consists of an array
containing six nodal numbers, ordered cyclically so as define an
orientation for the face; pointers to the top and bottom
tetrahedra, defined in accordance with the orientation of the
face, Fig.~\ref{assembledelements}a; pointers to the three edges
incident on the face; and a boolean variable indicative of
whether the face is interior or exterior, Fig.~\ref{codefacet}.
Specifically, the identify top (bottom) or positive (negative)
\ttet~ is that tetrahedron from which the nodes of the \tfac~
appear to be traversed counterclockwise (clockwise).  For \tfacs~
which are on the surface, the bottom \ttet~ is null. With a view
to fragmentation applications, we additional collect the
components of the unit normal; the critical traction for the
insertion of a cohesive element, and an index designating the
cohesive law to be used within the cohesive element.

\begin{figure}[H]
\begin{center}
\begin{code}

/*--------------------------------------------------------------*
 * Facet data structure                                         *
 *--------------------------------------------------------------*/
typedef struct facet
\{
  struct facet   *Link, *Rink;  /*{\it Links }*/
  int             N[6];         /*{\it Nodes }*/
  struct tetra   *T[2];         /*{\it Pointers to Tets }*/
  struct segment *S[3];         /*{\it Pointers to Segments }*/
  int             posi;         /*{\it 0 = internal, 1 = external }*/
  double          v[3];         /*{\it Components of the unit normal }*/
  double          sc;           /*{\it Traction limit value }*/
  int             cm;           /*{\it Cohesive material index }*/
\} Facet;

\end{code}
\vspace{0.1in}
\caption[]{\small Description of the \tfac~ data structure.}
\label{codefacet}
\end{center}
\end{figure}

The edge structure \tseg~ contains an array with the numbers of
its three nodes, Fig.~\ref{codesegment}, all of which are shared
with one or more adjacent \tfacs~ and with one or more adjacent
\ttets. The local sequential numbering of the nodes defines the
orientation of the edge. Owing to the variable environment of the
edges in the triangulation, the structure \tsegs~ additionally
contains arrays of pointers which need to be allocated
dynamically. These are the array of pointers to the tetrahedra
and faces adjacent to the edge. The dimension of these arrays is
also stored in the structure, as well as an additional boolean
variable which designates the edge as interior or exterior.

\begin{figure}[H]
\begin{center}
\begin{code}

/*--------------------------------------------------------------*
 * Segment data structure                                       *
 *--------------------------------------------------------------*/
typedef struct segment
\{
  struct segment *Link, *Rink;  /*{\it Links }*/
  int             N[3];         /*{\it Nodes }*/
  int             nt;           /*{\it No. adj tets }*/
  struct tetra  **T;            /*{\it Pointers to the first Tet }*/
  int             nf;           /*{\it No. adj Facets }*/
  struct facet  **F;            /*{\it Pointer to the first Facet }*/
  int             posi;         /*{\it 0 = internal, 1 = external }*/
\} Segment;

\end{code}
\vspace{0.1in}
\caption[]{\small Description of the \tseg~ data structure.}
\label{codesegment}
\end{center}
\end{figure}

A C procedure for setting up the structures just described from a
conventional finite-element connectivity array is shown in
Fig.~\ref{codeloadds}. As the connectivity table is traversed,
each element is added to the linked list of \ttets. The
corresponding six \tsegs~ of each \ttets~ are identified, and the
linked list of the \tsegs~ and the list of incident \ttets~ of the
\tsegs~ are updated. Likewise, the four \tfacs~ of each tetrahedra
are identified, and the linked list of the \tfacs~ and the list of
incident \ttets~ of the \tfacs~ are updated.
\begin{figure}[H]
\begin{center}
\begin{code}

int sc[6][4] = \{\{0,4,1\},\{1,5,2\},\{2,6,0\},   /* {\it segments topology} */ 
                 \{0,7,3\},\{1,8,3\},\{2,9,3\}\};
int fc[4][6] = \{\{0,1,2,4,5,6\},\{0,3,1,7,8,4\}, /* {\it faces topology} */ 
               \{1,3,2,8,9,5\},\{0,2,3,6,9,7\}\};
int p1[3] = \{1,2,0\};
/*--------------------------------------------------------------*
 * void CreateDataStructures                                    *
 *--------------------------------------------------------------*/
\{
  Segment  *S[6], *G;
  Facet    *F[4], *R;
  Tetra    *T,    *Ti;
  int       ie, ic, i, j;
  int       n[10];
  TetraList   = NULL;    /* {\it Initialize the Tetrahedra list } */
  SegmentList = NULL;    /* {\it Initialize the Segments list } */
  FacetList   = NULL;    /* {\it Initialize the Facets list } */
  for (ie = 0; ie < TetraElementNumber; ie++) \{
                                   /* {\it Add Tetrahedron} */
    for (j = 0; j < 10; j++) n[j] = connectivity[ie*nodes_element+j];
    TetraList = T = AddTetra (TetraList, n[0], n[1], n[2], n[3], n[4],
                              n[5], n[6], n[7], n[8], n[9], ie);
    for (ic = 0; ic < 6; ic++) \{  /* {\it Add Segments} */
      S[ic] = SearchInSegments (T, n[sc[ic][0]], n[sc[ic][1]],
                                   n[sc[ic][2]]);
      TetraList->S[ic] = S[ic];
    \}
    for (ic = 0; ic < 4; ic++) \{  /* {\it Add Facets} */
      F[ic] = SearchInFacets (T, n[fc[ic][0]], n[fc[ic][1]],
                                 n[fc[ic][2]], n[fc[ic][3]],
                                 n[fc[ic][4]], n[fc[ic][5]]);
      for (j = 0; j < 3; j++) \{   /* {\it Check Segments and Facets} */
         for (i = 0; i < 6; i++) \{
            if ((S[i]->N[0] == F[ic]->N[j] &&
                 S[i]->N[2] == F[ic]->N[p1[j]]) ||
                (S[i]->N[2] == F[ic]->N[j] &&
                 S[i]->N[0] == F[ic]->N[p1[j]])) \{
            F[ic]->S[j] = S[i];
            SegmentFacet (S[i], F[ic]);
          \}
        \}
      \}
      /* {\it update the incident Facet} */
      TetraList->F[ic] = F[ic];
    \}
  \}
\end{code}
\end{center}
\end{figure}

\begin{figure}[H]
\begin{center}
\begin{code}
/* {\it Complete the Tet list with adj Tets} */
  for (T = TetraList; T != NULL; T = T->Rink) \{
    for (i = 0; i < 4; i++) \{
      R = T->F[i];
      if (R->T[0] == T) T->T[i] = R->T[1];
      if (R->T[1] == T) T->T[i] = R->T[0];
    \}
  \}
/* {\it Complete the Facets list defining if inside or outside} */
  for (R = FacetList; R != NULL; R = R->Rink) \{
    if (R->T[0] == NULL) R->posi = 1;
    if (R->T[1] == NULL) R->posi = 1;
  \}
/* {\it Complete the Segment list defining if inside or outside} */
  for (G = SegmentList; G != NULL; G = G->Rink) \{
    for (i = 0; i < G->nf; i++) \{
      if (G->F[i]->posi == 1) G->posi = 1;
    \}
  \}
  return;
\}
\end{code}
\vspace{0.1in} \caption[]{\small This procedure fills the data
structures described in Fig.~\ref{codetetra}-\ref{codesegment}. 
A new \tseg~ or \tfac~ insertion follows from
the failure of a search in the corresponding linked lists.
Subroutines are described in Fig.~\ref{sub1}-\ref{sub4}.}
\label{codeloadds}
\end{center}
\end{figure}

\begin{figure}[H]
\begin{center}
\begin{code}
/*----------------------------------------------------------------*
 * Tetra *AddTetra                                                *
 *----------------------------------------------------------------*/
Tetra
*AddTetra (Tetra *beg, int n1, int n2, int n3, int n4, int n5,
	   int n6, int n7, int n8, int n9, int n10, int ie)
\{
  Tetra *T;      /* {\it Add a Tet to the linked list} */
  T       = calloc (1, sizeof(Tetra));
  T->el   = ie + 1;
  T->N[0] = n1;  T->N[1] = n2;
  T->N[2] = n3;  T->N[3] = n4;
  T->N[4] = n5;  T->N[5] = n6;
  T->N[6] = n7;  T->N[7] = n8;
  T->N[8] = n9;  T->N[9] = n10;
  T->C[0] = 0;   T->C[1] = 0;  T->C[2] = 0;   T->C[3] = 0;
  T->Link = NULL;
  T->Rink = beg;
  if (beg != NULL) beg->Link = T;
  return T;
\}
/*----------------------------------------------------------------*
 * Segment *SearchInSegments                                      *
 *----------------------------------------------------------------*/
Segment
*SearchInSegment (Tetra *T, int n1, int n2, int n3)
\{
  Segment  *S;    /* {\it Search for a Segment in SegmentList} */
  int       nt;
  for (S = SegmentList; S != NULL; S = S->Rink) \{
    if ((n1 == S->N[0] && n3 == S->N[2]) ||
        (n1 == S->N[2] && n3 == S->N[0])) \{
      nt = S->nt;
      nt++;
      S->T = realloc (S->T, nt * sizeof(Tetra *));
      S->T[nt - 1] = T;
      S->nt = nt;
      return S;
    \}
  \}
  SegmentList = S = AddSegment (SegmentList, n1, n2, n3, T);
  return S;
\}
\end{code}
\vspace{0.1in}
\caption[]{\small Insertion of a new \ttet~ and check of
the existence of a \tseg~ in the corresponding linked lists.}
\label{sub1}
\end{center}
\end{figure}

\begin{figure}[H]
\begin{center}
\begin{code}
/*----------------------------------------------------------------*
 * Segment *AddSegment                                            *
 *----------------------------------------------------------------*/
Segment
*AddSegment (Segment *beg, int n1, int n2, int n3, Tetra *T)
\{
  Segment *S;     /* {\it Add a Segment to the linked list} */
  S       = calloc (1, sizeof(Segment));
  S->posi = 0;
  S->N[0] = n1;  S->N[1] = n2;  S->N[2] = n3;
  S->nt   = 1;
  S->T    = calloc (1, sizeof(Tetra *));
  S->T[0] = T;
  S->nf   = 0;
  S->F    = calloc (1, sizeof(Facet *));
  S->Rink = beg;
  S->Link = NULL;
  if (beg != NULL) beg->Link = S;
  return S;
\}
/*----------------------------------------------------------------*
 * Segment *SearchInFacets                                        *
 *----------------------------------------------------------------*/
Facet
*SearchInFacets (Tetra *T, int n1, int n2, int n3,
                 int n4, int n5, int n6)
\{
  Facet *F;      /* {\it Search for a Facet in the FacetList} */
  for (F = FacetList; F != NULL; F = F->Rink) \{
    if ((n1 == F->N[0] || n1 == F->N[1] || n1 == F->N[2]) &&
        (n2 == F->N[0] || n2 == F->N[1] || n2 == F->N[2]) &&
        (n3 == F->N[0] || n3 == F->N[1] || n3 == F->N[2])) \{
      F->T[1] = T;
      return F;
    \}
  \}
  FacetList = F = AddFacet (FacetList, n1, n2, n3, n4, n5, n6, T);
  return F;
\}
\end{code}
\vspace{0.1in}
\caption[]{\small Insertion of a new \tseg~ and check of
the existence of a \tfac~ in the corresponding linked lists.}
\label{sub2}
\end{center}
\end{figure}

\begin{figure}[H]
\begin{center}
\begin{code}
/*----------------------------------------------------------------*
 * Facet *AddFacet                                                *
 *----------------------------------------------------------------*/
Facet
*AddFacet (Facet *beg, int n1, int n2, int n3,
           int n4, int n5, int n6, Tetra *T)
\{
  Facet *F;     /* {\it Add a Facet to the linked list} */
  F       = calloc (1, sizeof(Facet));
  F->posi = 0;
  F->N[0] = n1;  F->N[1] = n2;  F->N[2] = n3;
  F->N[3] = n4;  F->N[4] = n5;  F->N[5] = n6;
  F->T[0] = T;
  F->Rink = beg;
  F->Link = NULL;
  if (beg != NULL) beg->Link = F;
  return F;
\}
/*----------------------------------------------------------------*
 * void *SegmentFacet                                             *
 *----------------------------------------------------------------*/
void
SegmentFacet (Segment *S, Facet *F)
\{
  int nf = S->nf;
  int j;
  for (j = 0; j < nf; j++) if (S->F[j] == F) return;
  AddFaceToSegment (S, F);
  return;
\}
/*----------------------------------------------------------------*
 * void *AddFaceToSegment                                         *
 *----------------------------------------------------------------*/
void
AddFaceToSegment (Segment *S, Facet *F)
\{
  int nf = S->nf;
  nf++;
  S->F  = realloc (S->F, nf*sizeof(Facet));
  S->F[nf - 1] = F;
  S->nf = nf;
  return;
\}
\end{code}
\vspace{0.1in}
\caption[]{\small Insertion of a new \tfac~ in the corresponding
linked list, update of the list of \tfac~ in the \tseg~ data
structure.}
\label{sub4}
\end{center}
\end{figure}

\section{Fragmentation}
\label{fragmentation}

\begin{figure}
\begin{center}
\subfigure[]{\epsfig{figure=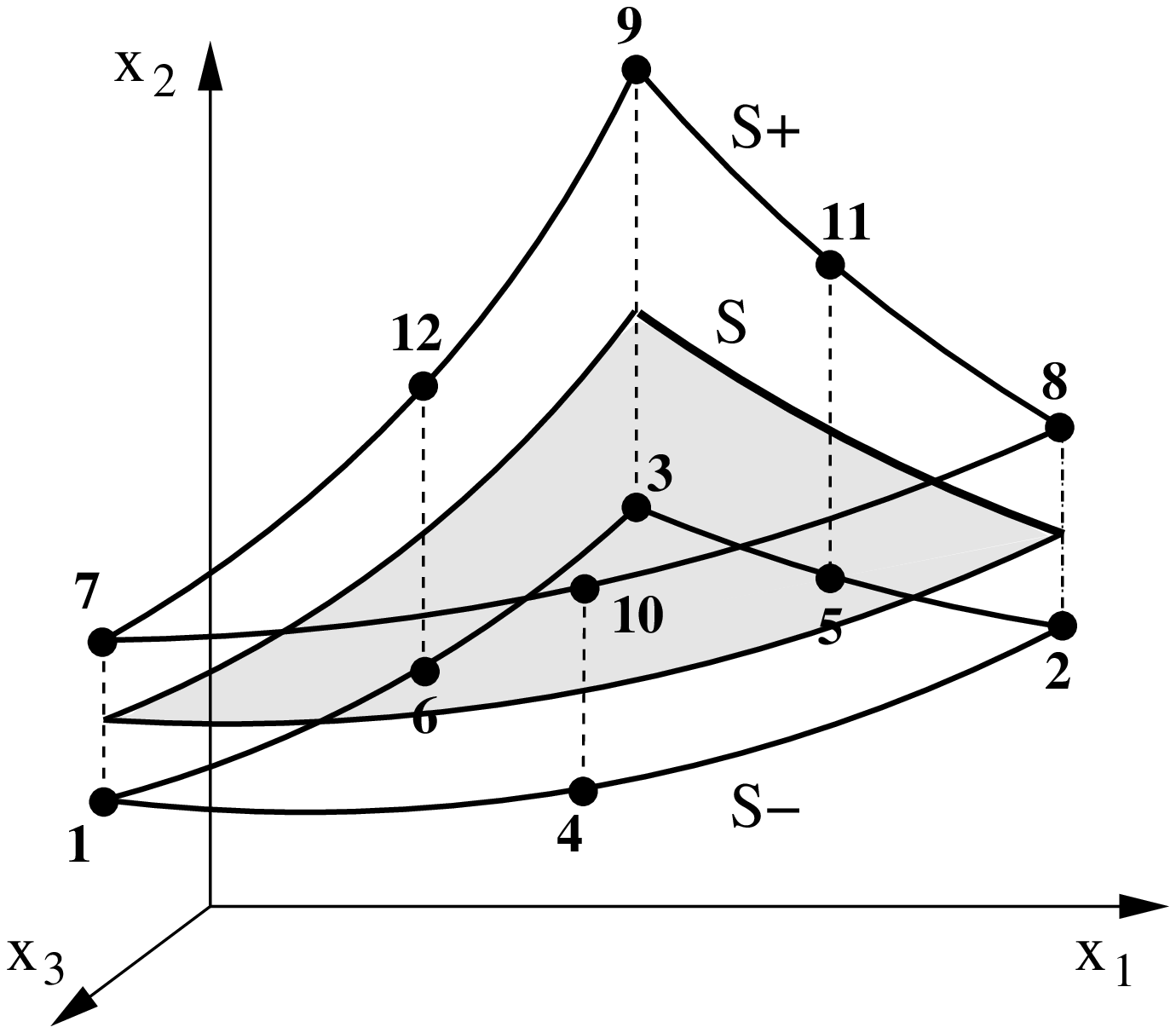,height=2.3in}}
\hspace{0.6in}
\subfigure[]{\epsfig{figure=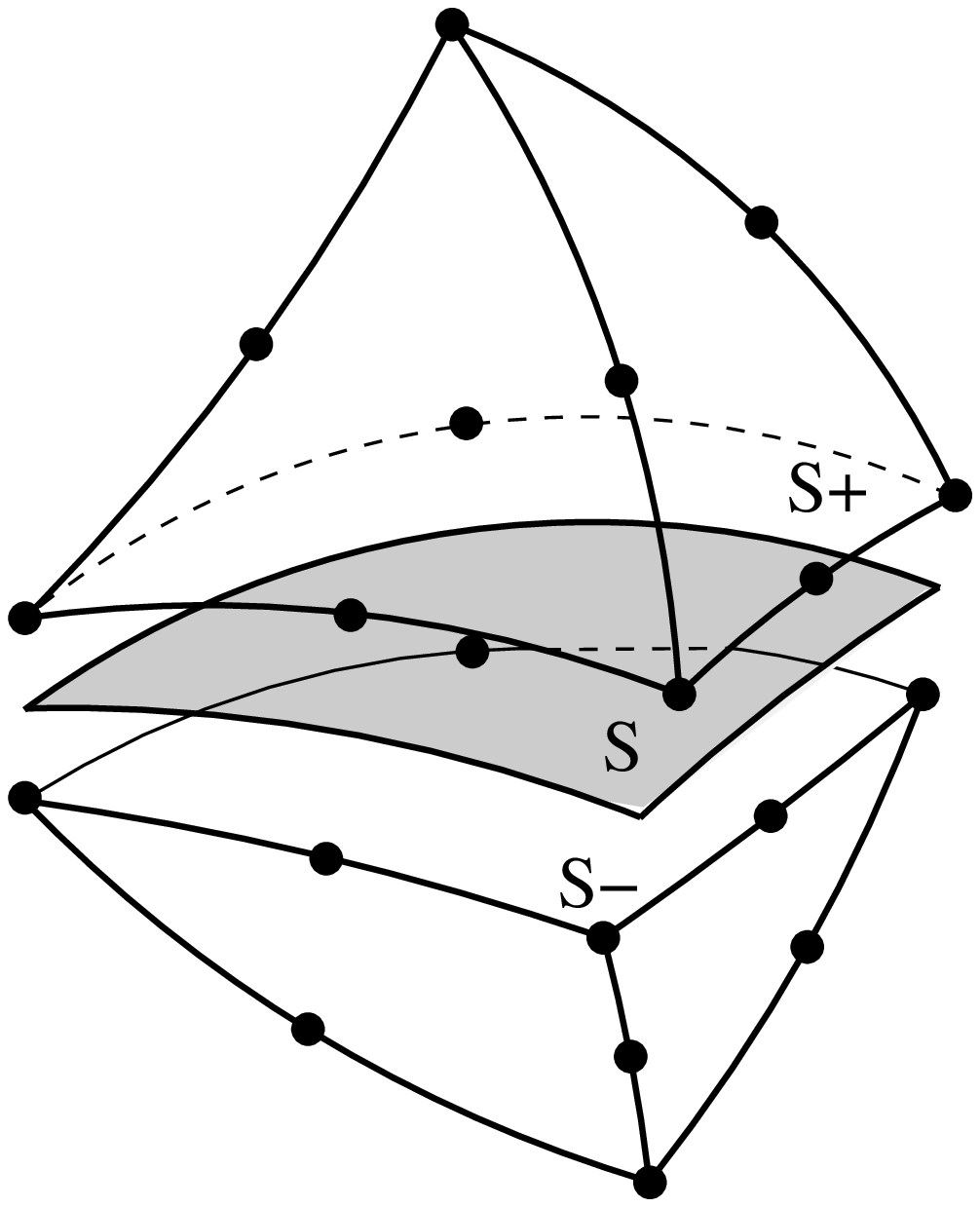,height=2.3in}}
\caption[]{\small (a) Geometry and connectivities of 12-nodes
cohesive element; (b) Assembly of 12-node triangular cohesive
element and two 10-node tetrahedral elements.} \label{cohtet}
\end{center}
\end{figure}

Ortiz and Pandolfi \cite{ortiz:1999} developed a class of
three-dimensional cohesive elements consisting of two six-node
triangular facets (Fig.~\ref{cohtet}a). The opening displacements
are described by quadratic interpolation within the element. The
element is fully compatible with --and may be used to bridge--
pairs of ten-node tetrahedral elements (Fig.~\ref{cohtet}b). The
elements are endowed with full finite-deformation kinematics and,
in particular, are exactly invariant with respect to superposed
rigid body translations and rotations.

We are here specially concerned with dynamic fragmentation,
although static applications may be treated similarly. The
analysis proceeds incrementally in time, e.~g., by explicit
dynamics.  Following Camacho and Ortiz \cite{camacho:1996},
cohesive elements are introduced adaptively at element interfaces
as required by a fracture --or spall-- criterion. For instance,
fracture may be supposed to initiate at a previously coherent
element interface when a suitably defined {\it effective
traction} attains a critical value \cite{camacho:1996,
deandres:1999, ortiz:1999}. When the fracture criterion is met at
an element interface, a cohesive element is inserted, leading to
the creation of new surface.  In this manner, the shape and
location of successive crack fronts is itself an outcome of the
calculations.

Next we discussed how the data structures defined in the
preceding section may be used to support fragmentation
simulations of the type just described. In particular, we
specifically address the issue of how to update the data
structures in response to fragmentation. The update procedure
consists of two basic operations:

\begin{enumerate}

\item Selection of interior faces (\tfacs) for the insertion of
new cohesive elements.

\item Updating the data structures based on the selected 
\tfacs.

\end{enumerate}

\noindent We proceed to discuss these two steps in turn.

\subsection{Selection of faces to be fractured}
\label{Selction}

\begin{figure}
\begin{center}
\epsfig{file=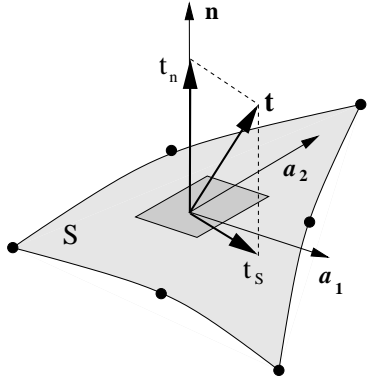,width=2in} \caption[]{\small
Normal and tangential components of the traction acting on an
interface.} \label{traction}
\end{center}
\end{figure}

The selection of an interior face for the insertion of a cohesive
element is based on the attainment of a suitable fracture
criterion. The specific form of this criterion depends on the type
of cohesive model used in calculations. For definiteness, we
consider the class of cohesive laws proposed by
\cite{camacho:1996, deandres:1999, ortiz:1999}, which are based
on the introduction of an effective opening displacement and the
corresponding work-conjugate effective traction:
\begin{equation}
t_{eff} = \sqrt{t_n^2 + \beta^{-2} |\bt_s|^2}
\label{EffectiveTraction}
\end{equation}
where $\beta$ is a tension-shear coupling parameter, $t_n$
denotes the traction component normal to the \tfac, and $\bt_s$
is the corresponding tangential traction, Fig.~\ref{traction}.
If, in addition, cohesive surfaces are assumed to be rigid, or
perfectly coherent, below a certain critical traction $\sigma_c$,
or spall strength, then the appropriate form of the fracture
criterion is:
\begin{equation}
t_{eff} \ge \sigma_c \label{SigmaCritical}
\end{equation}
This condition is checked for each internal face at the
conclusion of a prespecified number of time steps in the
calculations, and the faces where the criterion is met are
flagged for subsequent processing. Evidently, criterion
(\ref{SigmaCritical}), which arises directly from the mechanics of
cohesive fracture, drives the evolution of the geometrical
description of the model. This connection exemplifies the tight
coupling between mechanics and geometry which is characteristic of
fragmentation simulations.

\subsection{Data-structure update}
\label{meshupdate}

\begin{figure}[H]
\begin{center}
\epsfig{figure=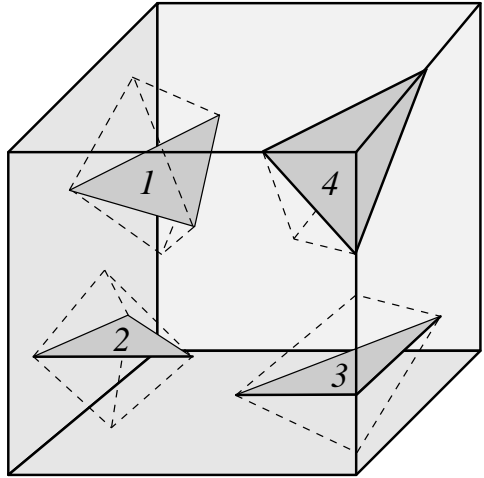,height=2.5in}
\vspace{0.1in} \caption[]{\small Classification of cases
according to whether the fractured \tfac~ has zero (case 1), one
(case 2), two (case 3) or three (case 4) \tsegs~ on the boundary.}
\label{openingcase}
\end{center}
\end{figure}

\begin{figure}[H]
\begin{center}
\subfigure[]{\epsfig{figure=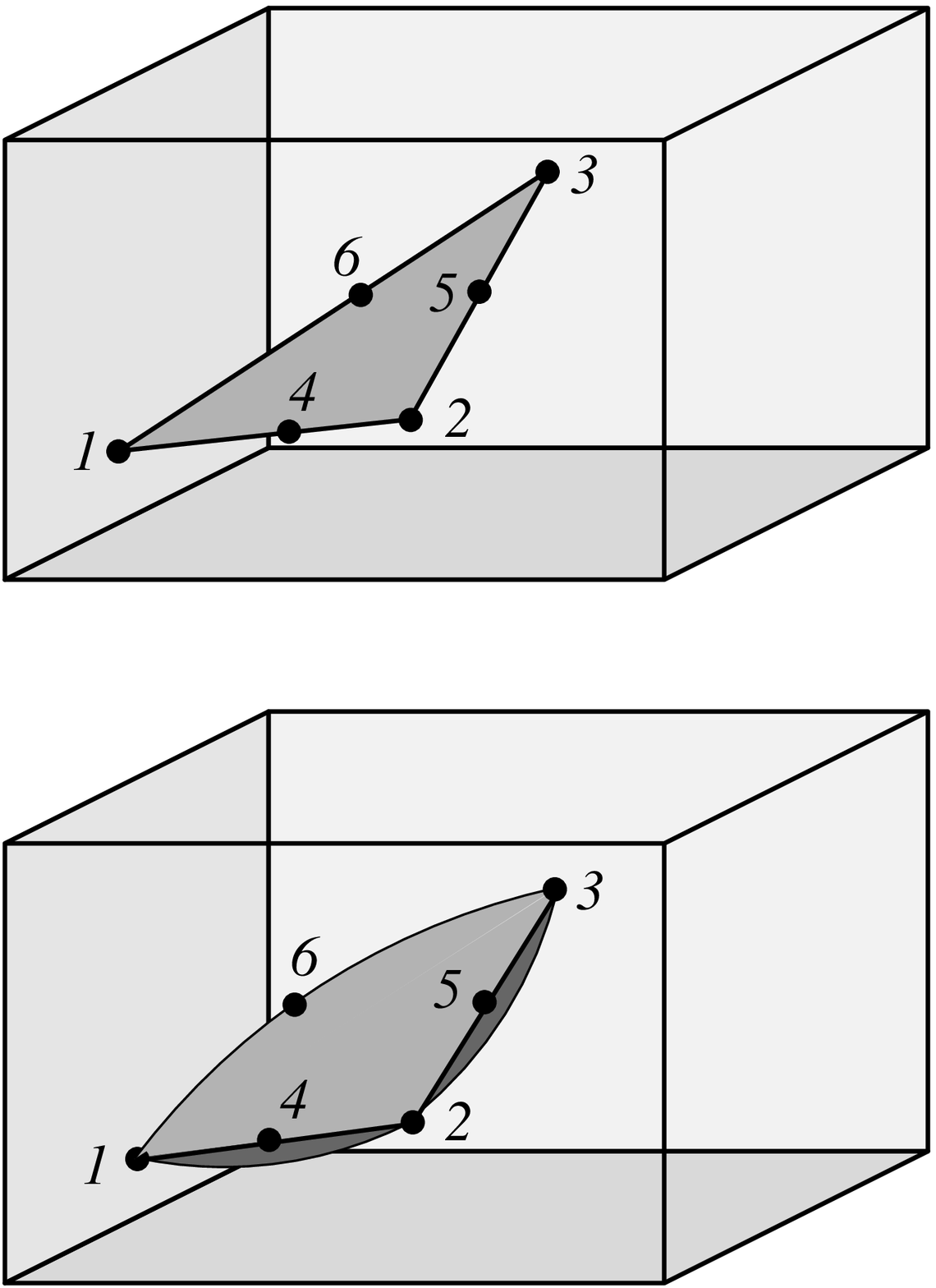,height=2.5in}}
\hspace{0.5in}
\subfigure[]{\epsfig{figure=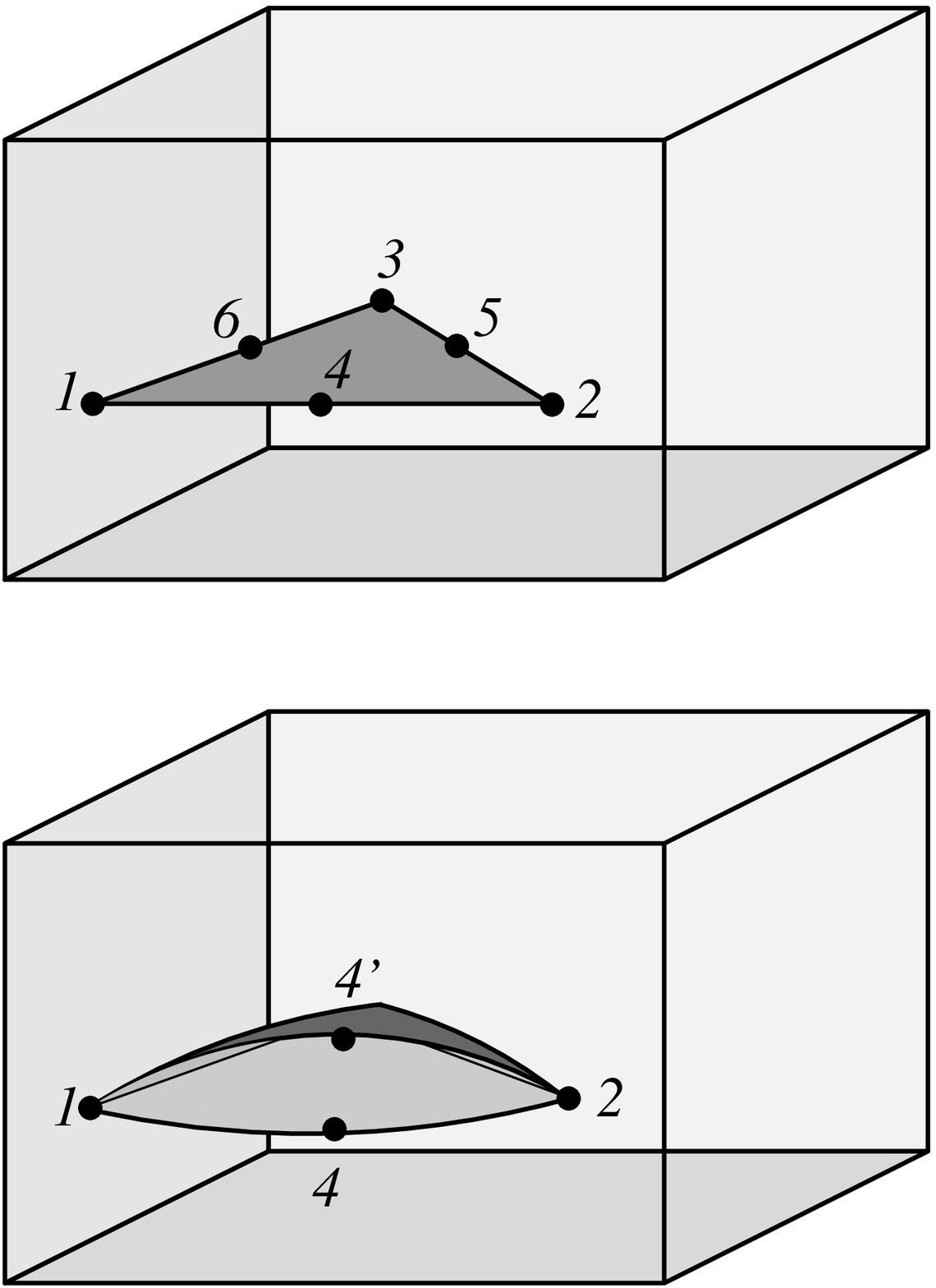,height=2.5in}}
\subfigure[]{\epsfig{figure=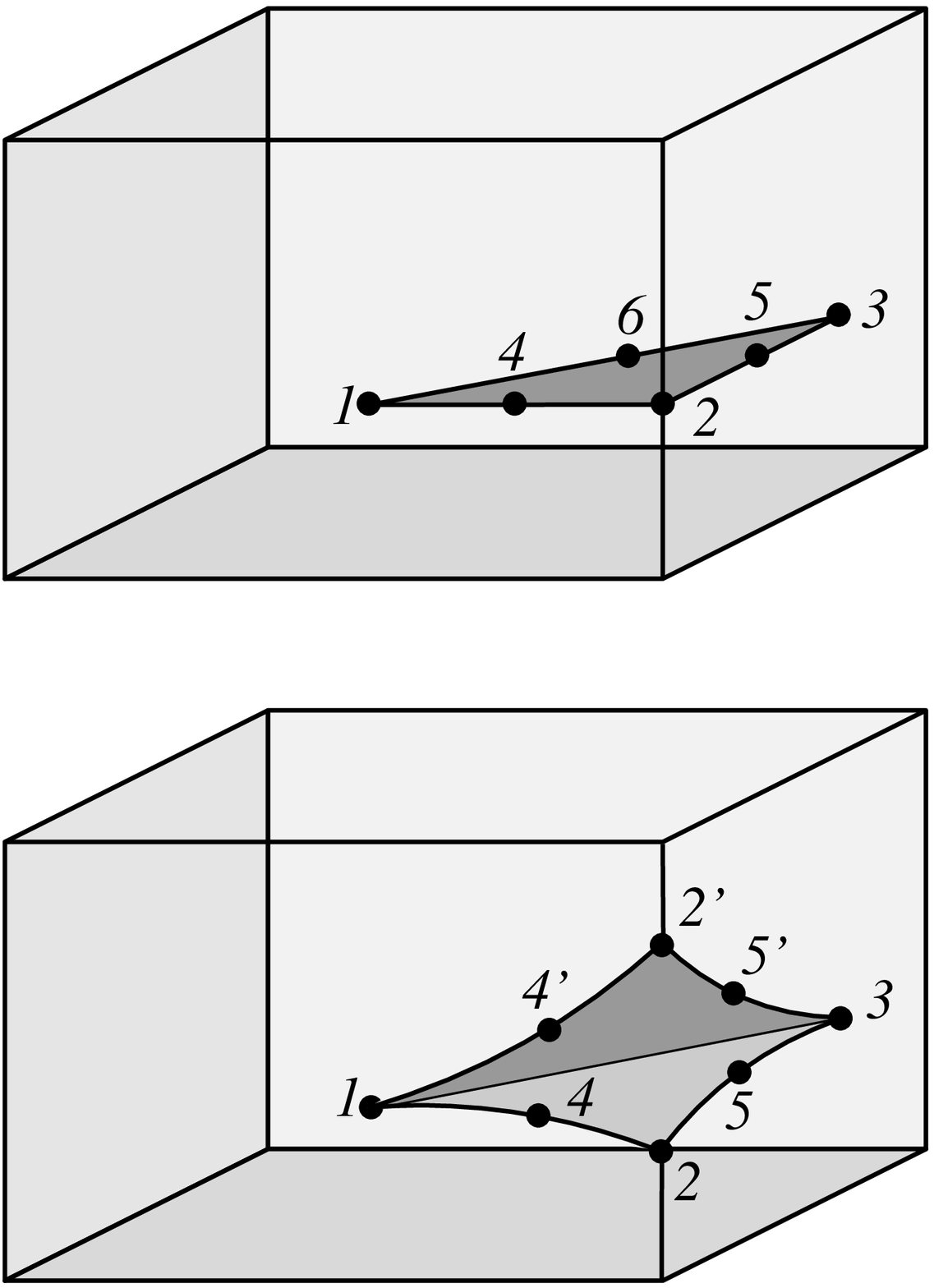,height=2.5in}}
\hspace{0.5in}
\subfigure[]{\epsfig{figure=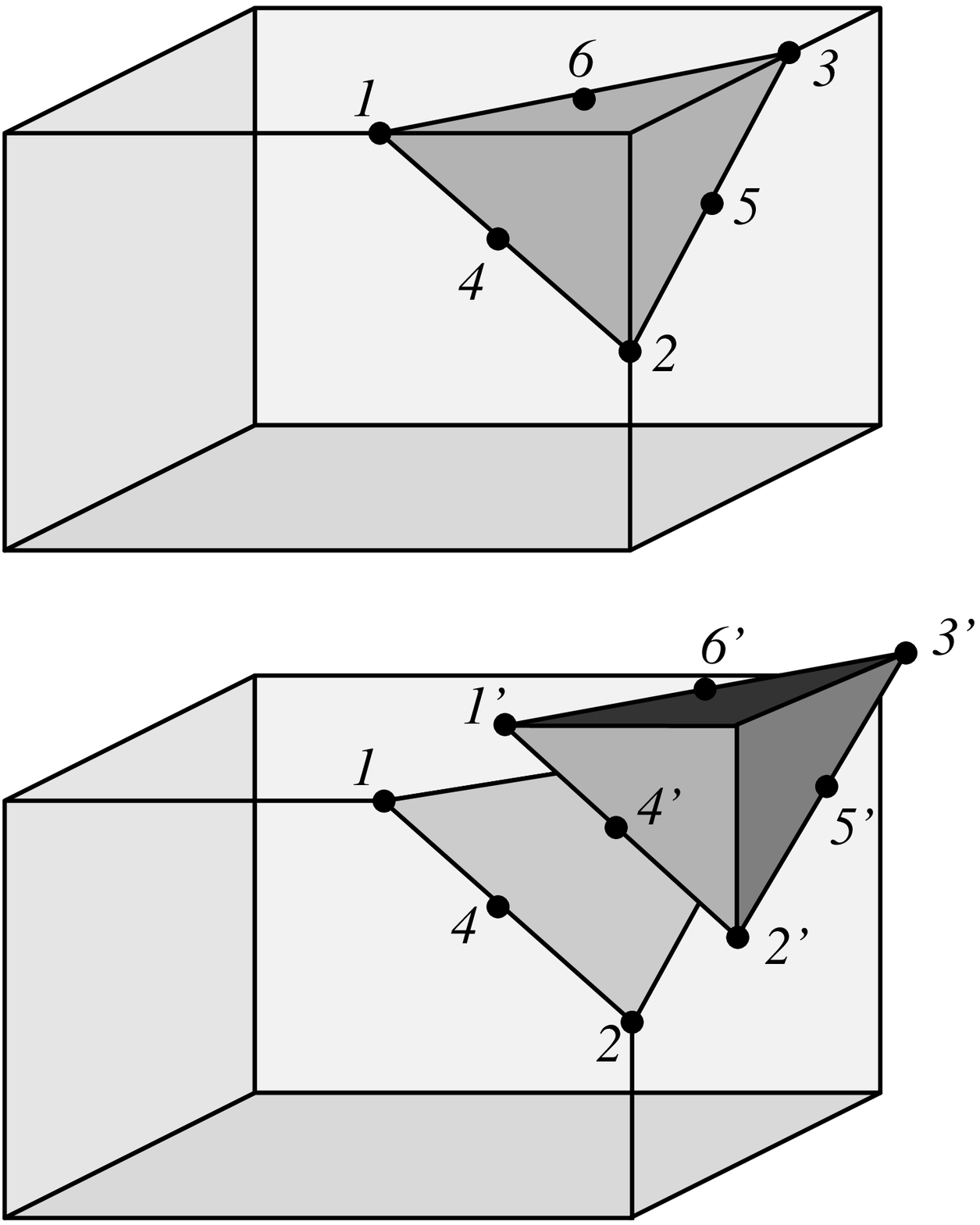,height=2.5in}}
\vspace{0.1in} \caption[]{\small Topological changes induced by
the fracturing of a \tfac. The primed numbers indicate the new
inserted nodes. (a) Case 1: no \tsegs~ on the boundary. No action; 
(b) Case 2: one \tseg~ on the boundary. The mid-side node is 
duplicated. (c) Case 3: two \tsegs~ on the boundary. The
corner node is duplicated and the mid-side nodes are duplicated;
(d) Case 4: three \tsegs~ on the boundary. All six nodes are
duplicated.} \label{fourcases}
\end{center}
\end{figure}

The operations to perform in order to process a fractured face
depend critically on its position with respect to the external
boundaries of the domain. Four main cases may be identified
depending on whether the fractured \tfac~ has zero, one, two or
three \tsegs~ resting on the boundary, Fig.~\ref{openingcase}.
This information is supplied by the flag {\it posi} of each \tseg~
contained in the \tfac. In all cases the fractured \tfac~ is
duplicated and a new \tfac~ is added to the corresponding linked
list. The remaining operations to be performed are:
\begin{enumerate}
\item No \tsegs~ are on the boundary: No further operations,
Fig.~\ref{fourcases}a.
\item One \tseg~ is on the boundary: the \tseg~ is
duplicated by doubling the mid-side node, Fig.~\ref{fourcases}b.
\item Two \tsegs~ are on the boundary: the \tsegs~ are
duplicated by doubling the mid-side nodes; the corner node is
duplicated when it represents the sole remaining connection
between the top and bottom \ttets, Fig.~\ref{fourcases}c.
\item Three \tsegs~ are on the boundary: the \tsegs~ are
duplicated by doubling the mid-side nodes; a corner node is
duplicated when it represents the sole remaining connection
between the top and bottom \ttets, Fig.~\ref{fourcases}d. The new
node is added to the top elements of the fractured \tfac.
\end{enumerate}
The commented C code reported in listing~\ref{main} gives a
detailed account of the operations to be performed in order to
update the data structures and the connectivities. Steps referred
to in the code are graphically displayed in
Figs.~\ref{step1}-\ref{step8}.

\begin{figure}[H]
\begin{center}
\begin{code}
/*--------------------------------------------------------------*
 * void CreateSurface                                           *
 *--------------------------------------------------------------*/
void
CreateSurface (int cohmate, Facet *facet)
\{
  Tetra   *TEB, *TET;   /* {\it Top and Bottom tetrahedra} */
  int      open[3];     /* {\it Boundary sides counters} */
  Segment *S[3];        /* {\it Facet Segments} */
  Facet   *facetnew;    /* {\it New Facet} */
  int      i, j;
                        /*{\it Get the six nodes on the Facet }*/
  for (i = 0; i < 6; i++) node[i] = nold[i] = facet->N[i];
  TET  = facet->T[0];  
  TEB  = facet->T[1];
  for (i = 0; i < 3; i++) \{
    S[i] = facet->S[i];
    open[i] = S[i]->posi;
  \}                               /*{\it add a new facet }*/
  FacetList = facetnew = AddFacet (FacetList, node[0], node[2], node[1],
                                   node[5], node[4], node[3], TET);
  UpdateNewFacet (facet, facetnew, TEB, TET, cohmate);
  for (i = 0; i < 3; i++) \{      /*{\it  Check the midside node }*/
    if (open[i] == 0) continue;
    nodes++;
    node[i + 3] = nodes;
    UpdateMidNodes (TET, S[i], node[i + 3], cohmate);
    DefineNodeVectors (node[i + 3], nold[i + 3]);
  \}                              /*{\it  Check the corner node }*/
  counted = calloc (elements + 1, sizeof(int));
  for (i = 0; i < 3; i++) \{
    if (open[i] == 0 || open[perm[i]] == 0) continue;
    for (j = 1; j <= TetraElementNumber; j++) counted[j] = 0;
    if (CornerToOpen (TET, TEB, nold[i]) == 0) continue;
    nodes++;
    node[i] = nodes;
    UpdateCornerNodes (TET, node[i], nold[i], cohmate);
    DefineNodeVectors (node[i], nold[i]);
  \}
  free (counted);
  if (cohmate > 0) AddCohesive (cohmate, trli);
  return;
\}
\end{code}
\vspace{0.1in} \caption[]{\small Main for the insertion of a new
surface. The subroutines are reported in
Figs.~\ref{step1}-\ref{step8}.} \label{main}
\end{center}
\end{figure}

\begin{figure}[H]
\begin{center}
\begin{code}
/*----------------------------------------------------------------*
 * void UpdateNewFacet                                            *
 *----------------------------------------------------------------*/
void
UpdateNewFacet (Facet *facet, Facet *facetnew,
                Tetra *TEB, Tetra *TET, int cohmate)
\{
  Segment *S[3];
  int      move[3] = \{2,1,0\};
  int      i;
/* {\it Update the old Facet }*/
  facet->T[1] = NULL;
  facet->posi = 1;
/* {\it Update the new Facet }*/
  facetnew->posi = 1;
/* {\it Add Segments to new Facet and new Facet to Segments }*/
  for (i = 0; i < 3; i++) \{
    S[i] = facet->S[i];
    S[i]->posi = 1;
    facetnew->S[move[i]] = S[i];
    S[i] = AddFaceToSegment (S[i], facetnew);
  \}
/* {\it Update the Facet and adjacency of the Tets }*/
  for (i = 0; i < 4; i++) \{
    if (TEB->F[i] == facet) \{
      TEB->T[i] = NULL;
      if (cohmate > 0) \{
        TEB->C[i] = elements + 1;
        TEB->L[i] = 0;
      \}
    \}
    if (TET->F[i] == facet) \{
      TET->F[i] = facetnew;
      TET->T[i] = NULL;
      if (cohmate > 0) \{
        TET->C[i] = elements + 1;
        TET->L[i] = 1;
      \}
    \}
  \}
  return;
\}
\end{code}
\vspace{0.1in}
\caption[]{\small Insertion of a new \tfac~ into the data structures.}
\label{step1}
\end{center}
\end{figure}

\begin{figure}[H]
\begin{center}
\begin{code}
/*----------------------------------------------------------------*
 * void UpdateMidNodes                                            *
 *----------------------------------------------------------------*/
void
UpdateMidNodes (Tetra *TET, Segment *SO, int nn, int cohmate)
\{
  Segment *SN;
  Tetra   *T;
  int      n1 = SO->N[0];
  int      n2 = SO->N[2];
  int      ot = SO->nt;
  int      of = SO->nf;

/* {\it Insert a new Segment }*/
  SegmentList = SN = AddSegment (SegmentList, n1, nn, n2, TET);
  SN->posi = 1;
  SN->nt   = 0;

/* {\it Transfer Tets from the old to the new Segment }*/
  SN->T = realloc (SN->T, ot*sizeof(Tetra));
  TetraInSegments (TET, SO, SN, cohmate);
  SN->T = realloc (SN->T, SN->nt*sizeof(Tetra));
  SO->T = realloc (SO->T, SO->nt*sizeof(Tetra));

/* {\it Transfer Facets between the two Segments }*/
  SN->F = realloc (SN->F, of*sizeof(Facet));
  FacetInSegments (SO, SN);
  SN->F = realloc (SN->F, SN->nf*sizeof(Facet));
  SO->F = realloc (SO->F, SO->nf*sizeof(Facet));
  return;
\}
\end{code}
\vspace{0.1in} \caption[]{\small Changes in the data structures
required by the duplication of a mid node. First, a new \tseg~ is
added (see Fig.~\ref{sub1}). Then, the list of the \ttets~
incident to the original \tseg~ is traversed and the \ttets~ are
moved from the old to the new \tseg~ (Fig.~\ref{step3}). Finally,
the list of the \tfacs~ incident to the original \tseg~ is
traversed and the \tfacs~ are moved from the old to the new \tseg~
(Fig.~\ref{step4}).} \label{step2}
\end{center}
\end{figure}

\begin{figure}[H]
\begin{center}
\begin{code}
/*----------------------------------------------------------------*
 * void TetraInSegments                                           *
 *----------------------------------------------------------------*/
void
TetraInSegments (Tetra *TET, Segment *SO, Segment *SN, int cohmate)
\{
  Tetra *T;
  int    el = (TET->el - 1)*nodes_element;
  int    no = SO->N[1];
  int    nn = SN->N[1];
  int    ot = SO->nt;
  int    nt = SN->nt;
  int    i, j;
  for (i = 4; i < 10; i++) \{ /* {\it Change the connectivity} */
    if (TET->N[i] != no) continue;
    TET->N[i] = connectivity[el + i] = nn;
    if (cohmate > 0) UpdateCohesive (TET, nn, no, 3);
  \}
  for (j = 0; j < 6; j++) \{  /* {\it Change the Segment in the Tet} */
    if (TET->S[j] != SO) continue;
    TET->S[j] = SN;
  \}
  for (i = 0; i < ot; i++) \{ /* {\it Move Tet from old to new Segment} */
    T = SO->T[i];
    if (T != TET) continue;
    SN->T[nt] = T;
    nt++;
    SO->T[i] = NULL;
  \}
  SN->nt = nt;
  for (j = i + 1; j < ot; j++, i++) SO->T[i] = SO->T[j];
  ot--;
  SO->nt = ot;
  for (i = 0; i < ot; i++) \{ /* {\it Look at adjacent Tets} */
    T = SO->T[i];
    for (j = 0; j < 4; j++) \{
      if (T->T[j] != TET) continue;
      TetraInSegments (T, SO, SN, cohmate);
      return;
    \}
  \}
  return;
\}

\end{code}
\vspace{0.1in}
\caption[]{\small Update the \ttets~ connected to the \tseg.}
\label{step3}
\end{center}
\end{figure}

\begin{figure}[H]
\begin{center}
\begin{code}
/*----------------------------------------------------------------*
 * void FacetInSegments                                           *
 *----------------------------------------------------------------*/
void
FacetInSegments (Segment *SO, Segment *SN)
\{
  Tetra *T;
  Facet *F;
  int    nn = SN->N[1];
  int    of = 0;
  int    nf = 0;
  int    i, j, k;
  for (i = 0; i < SN->nt; i++) \{ /* {\it Update connectivities} */
    T = SN->T[i];
    for (j = 0; j < 4; j++) \{
      F = T->F[j];
      for (k = 0; k < 3; k++) \{
        if (F->S[k] != SO) continue;
        F->S[k]     = SN;
        F->N[3 + k] = nn;
        SN->F[nf]   = F;
        nf++;
        break;
      \}
    \}
  \}
  SN->nf = nf;
  for (i = 0; i < SO->nt; i++) \{ /* {\it Update the incident Facets } */
    T = SO->T[i];
    for (j = 0; j < 4; j++) \{
      F = T->F[j];
      for (k = 0; k < 3; k++) \{
        if (F->S[k] != SO) continue;
        SO->F[of] = F;
        of++;
        break;
      \}
    \}
  \}
  SO->nf = of;
  return;
\}

\end{code}
\vspace{0.1in}
\caption[]{\small Update the \tfacs~ connected to the \tseg.}
\label{step4}
\end{center}
\end{figure}

\begin{figure}[H]
\begin{center}
\begin{code}
/*----------------------------------------------------------------*
 * void UpdateCohesive                                            *
 *----------------------------------------------------------------*/
void
UpdateCohesive (Tetra *T, int nn, int no, int add)
\{
  int ce, cc, i, j;
  for (i = 0; i < 4; i++) \{
    ce = T->C[i] - 1;
    if (ce < 0 || ce == elements) continue;
    cc = ce*nodes_element + T->L[i]*6 + add;
    for (j = 0; j < 3; j++) \{
      if (connectivity[cc + j] != no) continue;
      connectivity[cc + j] = nn;
    \}
  \}
  return;
\}
/*----------------------------------------------------------------*
 * void CornerToOpen                                              *
 *----------------------------------------------------------------*/
int
CornerToOpen (Tetra *TET, Tetra *TEB, int no)
\{
  Tetra *T;
  int    i, j, res;
  counted [TET->el] = 1;
  for (i = 0; i < 4; i++) \{
    T = TET->T[i];
    if (T == NULL) continue;
    for (j = 0; j < 4; j++) \{
      if (T->N[j] != no) continue;
      if (T == TEB) return 0;
      if (counted[T->el] > 0) continue;
      if (CornerToOpen (T, TEB, no) == 0) return 0;
    \}
  \}
  return 1;
\}
\end{code}
\vspace{0.1in} \caption[]{\small Update of the cohesive element
connectivity. The recursive subroutine {\tt CornerToOpen} checks
the existence of additional connections between the Top and the
Bottom \ttet~ before duplicating a corner node.} \label{step5}
\end{center}
\end{figure}

\begin{figure}[H]
\begin{center}
\begin{code}
/*----------------------------------------------------------------*
 * void UpdateCornerNodes                                         *
 *----------------------------------------------------------------*/
void
UpdateCornerNodes (Tetra *TET, int nn, int no, int cohmate)
\{
  Segment *S;
  Facet   *F;
  Tetra   *T;
  int      el = (TET->el - 1)*nodes_element;
  int      j, i;

  counted[TET->el] = 0;       /* {\it Flag the Tet as checked} */

  for (j = 0; j < 4; j++) \{   /* {\it Check Tet and connectivity} */
    if (TET->N[j] != no) continue;
    TET->N[j] = nn;
    connectivity[el + j] = nn;
    if (cohmate > 0) UpdateCohesive (TET, nn, no, 0);
    break;
  \}
  for (j = 0; j < 6; j++) \{   /* {\it Update Segments} */
     S = TET->S[j];
     if (S->N[0] == no) S->N[0] = nn;
     if (S->N[2] == no) S->N[2] = nn;
  \}
  for (i = 0; i < 4; i++) \{   /* {\it Update Facets} */
    F = TET->F[i];
    if (F->N[0] == no) F->N[0] = nn;
    if (F->N[1] == no) F->N[1] = nn;
    if (F->N[2] == no) F->N[2] = nn;
  \}
   for (i = 0; i < 4; i++) \{  /* {\it Next adjacency} */
    T = TET->T[i];
    if (T == NULL) continue;
    if (counted[T->el] == 0) continue;
    UpdateCornerNodes (T, nn, no, cohmate);
  \}
  return;
\}
\end{code}
\vspace{0.1in} \caption[]{\small Recursive subroutine to perform
changes in the data structures required by the duplication of a
corner node. First, the \ttet~ connectivity is update. Then the
incident \tsegs~ and \tfacs~ are updated. Finally, the adjacent
\ttets~ are checked.} \label{step7}
\end{center}
\end{figure}

\begin{figure}[H]
\begin{center}
\begin{code}
/*----------------------------------------------------------------*
 * void DefineNodeVectors                                         *
 *----------------------------------------------------------------*/
void
DefineNodeVectors (int node, int nold)
\{
  int nd = nodes*dof_node;
  int nn = (node - 1)*dof_node;
  int no = (nold - 1)*dof_node;
  int j;
  coordinates = realloc (coordinates, nd*REAL);
  for (j = 0; j < dof_node; j++) \{
    coordinates[nn + j] = coordinates[no + j];
  \}
  return;
\}

/*----------------------------------------------------------------*
 * void AddCohesive                                               *
 *----------------------------------------------------------------*/
void
AddCohesive (int cohmate, double trli)
\{
  int     k, i;
  k = nodes_element * elements;
  elements++;
  connectivity     = realloc (connectivity,     (k + 1)*sizeof(int));
  element_material = realloc (element_material, elements*sizeof(int));
  for (i = 0; i < 6; i++) \{
    connectivity[k + i    ] = nold[i];
    connectivity[k + i + 6] = node[i];
  \}
  element_material [elements - 1] = cohmate;
  return;
\}
\end{code}
\vspace{0.1in}
\caption[]{\small Reallocation of the nodal vectors
({\tt DefineNodeVectors}) and element vectors ({\tt
AddCohesive}). These subroutine must be completed with the
reallocation of all the node and element based vectors.}
\label{step8}
\end{center}
\end{figure}

\section{Examples of application}
\label{examples}

In this section we demonstrate the scope and versatility of the
procedures described above with the aid of two examples of
application. The fracture criterion (\ref{SigmaCritical}) is
adopted in order to determine the onset of fracture in a
\tfac~\cite{camacho:1996, deandres:1999, ortiz:1999}. In all
calculations, the material is modeled as nonlinear elastic
obeying a Neohookean constitutive law \cite{marsden:1983}. In particular,
full finite kinematics is taken into account.

\begin{figure}[H]
\begin{center}
\epsfig{figure=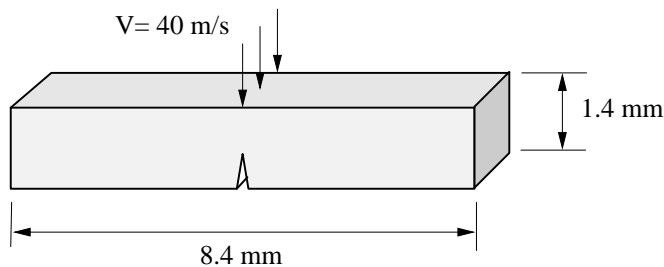,height=4.cm} \vspace{0.1in}
\caption[]{\small Three point bend test in PMMA specimen, loading
conditions. The impact is simulated imposing a uniform and constant
velocity $V$ along the central line. Owing to the high impact speed, 
the supports are not simulated.} \label{pmloading}
\end{center}
\end{figure}

The first application concerns the simulation of the dynamic
fragmentation of a three-point bend PMMA specimen with a sharp
precrack contained within its symmetry plane.  The central point
of the top side of the specimen is suddenly imparted a uniform
velocity 40~m/s at time $t=0$, and the velocity is held constant
thereafter. The length of the specimen is 8.4 mm, its width 1~mm
and its height 1.4~mm (Fig.~\ref{pmloading}). 

\begin{figure}[H]
\begin{center}
\psfig{figure=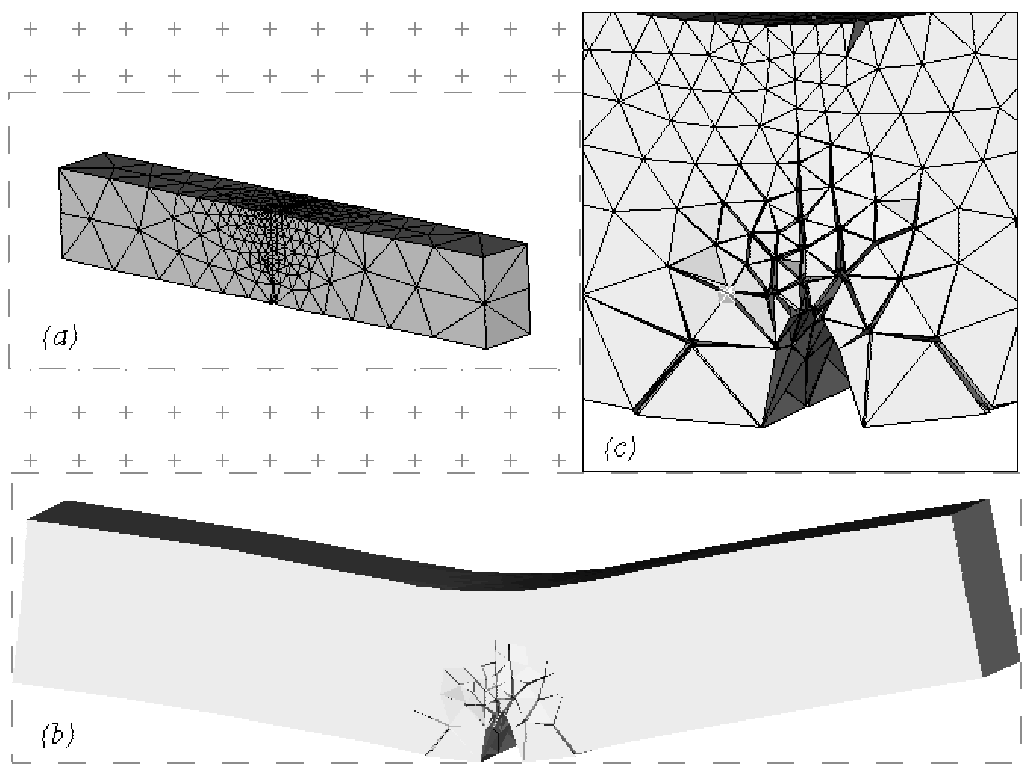,height=10cm} \vspace{0.1in}
\caption[]{\small Fragmentation algorithm applied to the three-point
bend dynamic test in PMMA: (a) initial mesh; (b) final
configuration; (c) detail of the fracture and fragmentation
pattern in the final configuration.} \label{pmma}
\end{center}
\end{figure}

\noindent The model is meshed into 4,260 ten-node tetrahedra and 6,420 
nodes. The mesh is finer in the central part and is gradually coarsened 
away from the crack (Fig.~\ref{pmma}a).

The material parameters employed in the calculations
are: specific fracture energy per unit area, or critical energy
release rate, $G_c = 210$ N/m; critical cohesive stress, or spall
strength, $\sigma_c = 100$ MPa; tension-coupling constant $\beta =
1$; Young's modulus $E = 3$ GPa; Poisson's ratio $\nu = 0.38$; and
mass density $\rho = 1180$ kg/m$^3$. The equations of motion are
integrated in time by recourse to Newmark's explicit algorithm
with parameters $\beta = 0$, $\gamma = 1/2$ \cite{belytschko:1983,
hughes:1983}. The time step used in the calculations is
$\triangle t = 1.7$ x $10^{-4} \mu$s.

Figs.~\ref{pmma}b and c show the computed fracture and
fragmentation pattern after 20~$\mu$s. The ability of the
approach to track the evolution of complex crack geometries is
noteworthy. As may be seen in the figure, the geometrical update
procedure effectively deals with the intricate geometrical and
topological transitions which result from crack branching, the
nucleation of surfaces and interior cracks, crack coalescence,
the detachment of fragments, and others.

\begin{figure}[H]
\begin{center}
\epsfig{figure=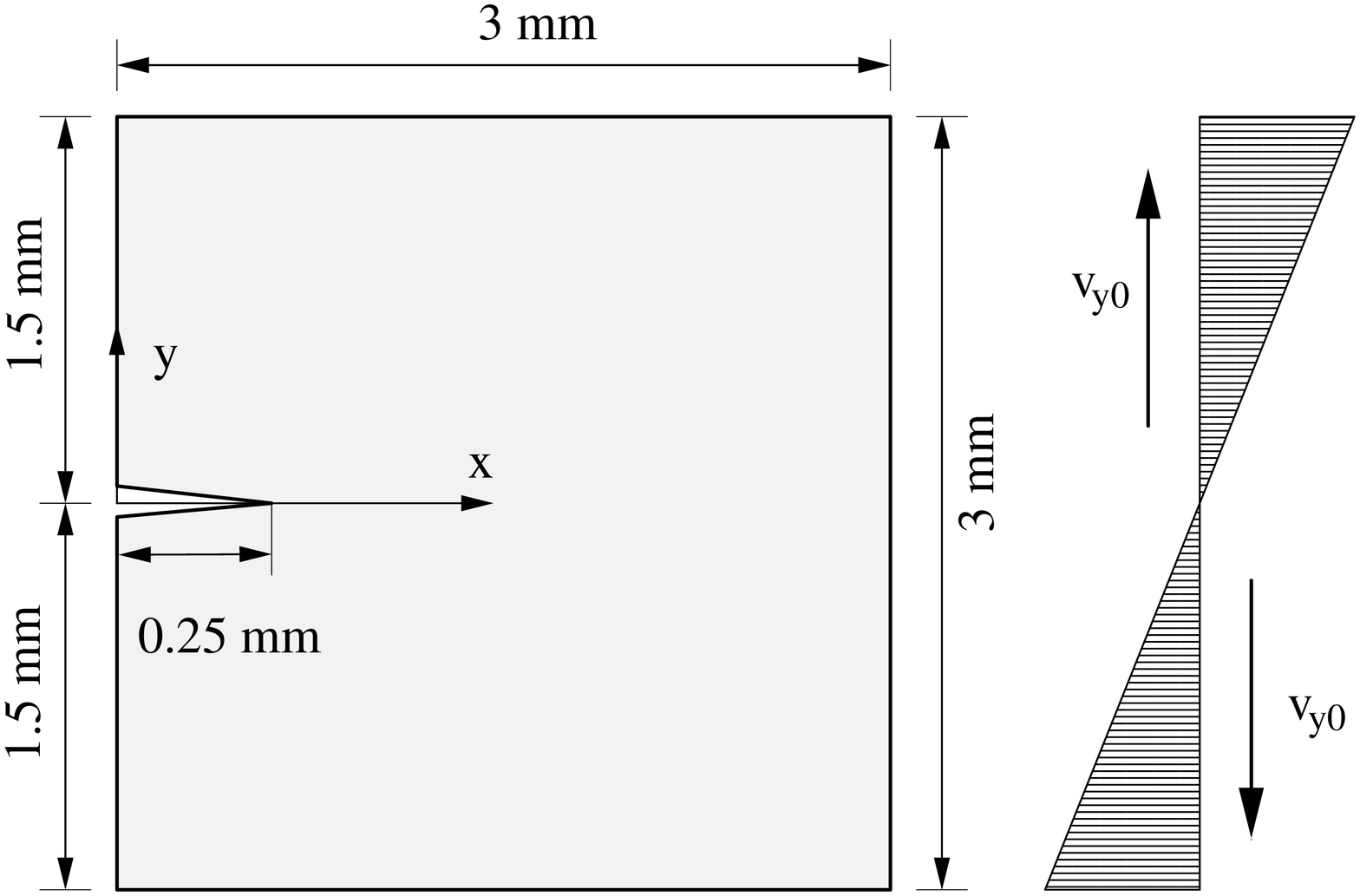,height=6cm} \vspace{0.1in}
\caption[]{\small Geometry of the PMMA square plate and
mode-I loading conditions.} \label{brloading}
\end{center}
\end{figure}

The second example of application concerns the simulation of
dynamic crack branching in a PMMA thin square plate. The plate is
3~mm in length and 0.3~mm in thickness. The specimen contains an
initial 0.25~mm sharp notch. Constant normal velocities, tending
to open the crack symmetrically in mode I, are prescribed on the
top and bottom edges of the specimen. The magnitude of the
prescribed velocities corresponds to a nominal strain rate of
0.002/$\mu$s. In addition, the initial velocity field is assumed
to be linear in the coordinate normal to the crack and to
correspond to a uniform rate of deformation of 0.002/$\mu$s
throughout the specimen, Fig.~\ref{brloading}.

\begin{figure}[H]
\begin{center}
\psfig{figure=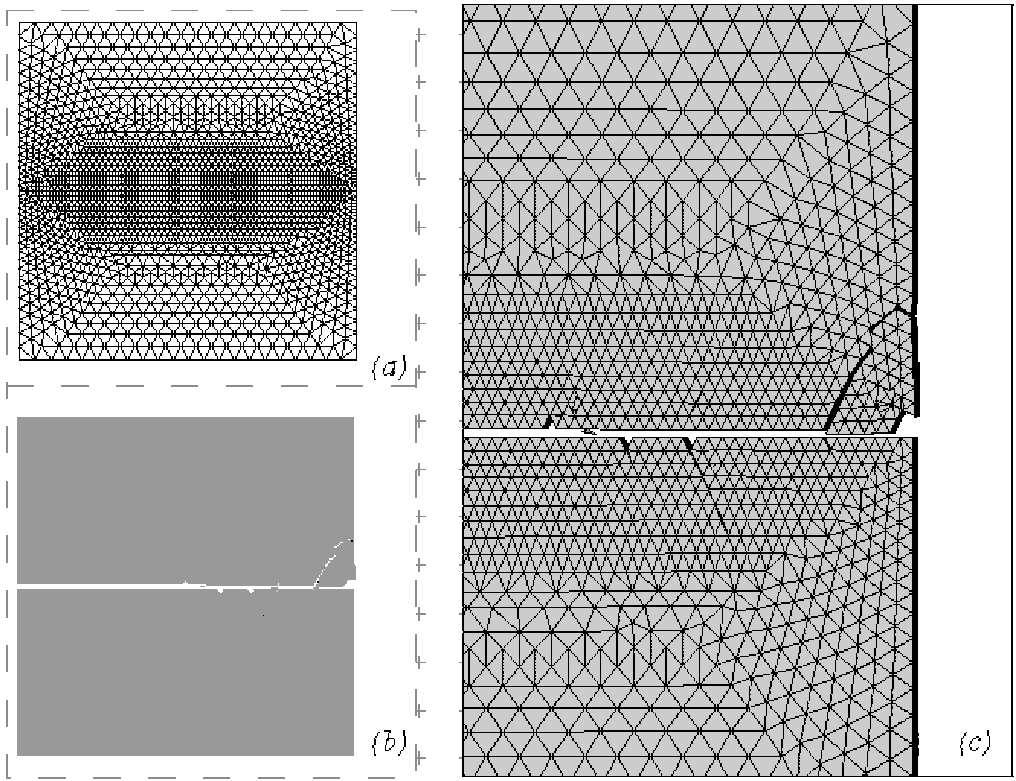,height=10.3cm} \vspace{0.1in}
\caption[]{\small Mode-I dynamic fracture test in PMMA at nominal
strain rate 0.002/$\mu$s. (a) Initial mesh and prescribed and initial
velocities; (b) final configuration; (c) detail of the branching
in the final configuration at time 5~$\mu$s.} \label{branching}
\end{center}
\end{figure}

\begin{figure}[H]
\begin{center}
\psfig{figure=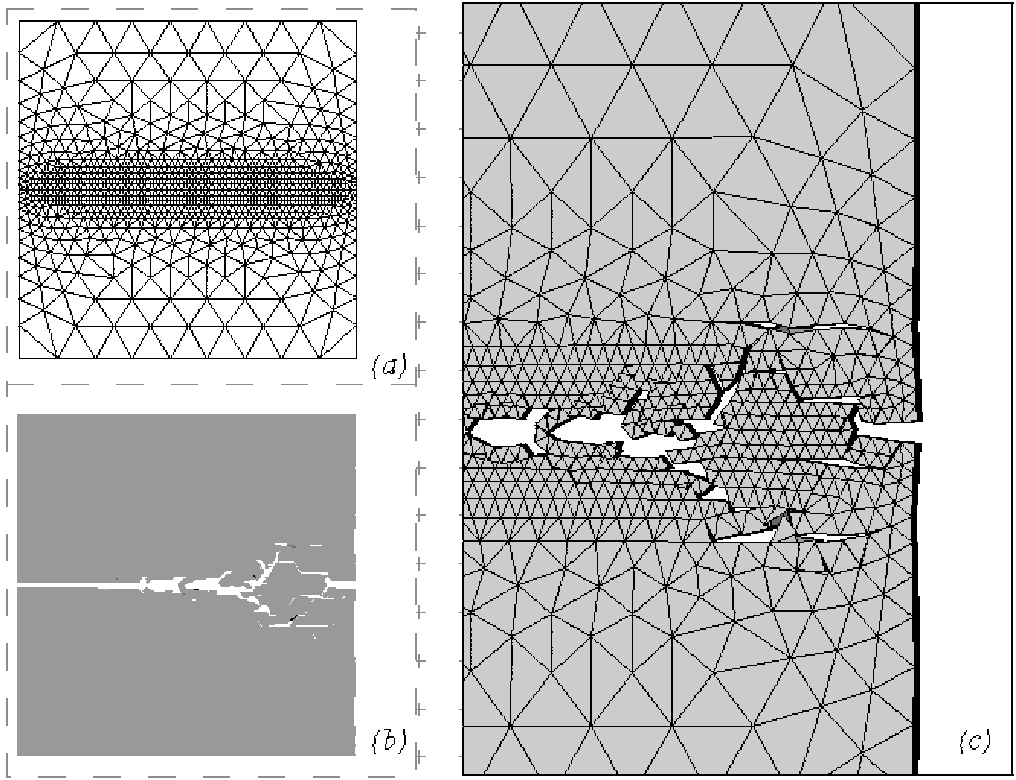,height=10.3cm}
\vspace{0.1in} \caption[]{\small Mode-I dynamic fracture test in
PMMA at a nominal strain rate of 0.01/$\mu$s. (a) Initial mesh; (b) final
configuration; (c) detail of the branching in the final
configuration at time 5~$\mu$s.} \label{branchingbig}
\end{center}
\end{figure}

The model is meshed into 14,319 ten-node tetrahedra and 25,936
nodes, Fig.~\ref{branching}a. The equations of motion 
are integrated using Newmark's
explicit algorithm with time step $\triangle t = 6$ x $10^-5 \mu$s.
The material parameters used in the  calculation are:
specific fracture energy per unit area $G_c = 176.15$ N/m;
critical cohesive stress $\sigma_c = 324$ MPa; tension-shear
coupling constant $\beta = 2$; Young's modulus $E = 3.29$ GPa;
Poisson's ratio $\nu = 0.35$; and mass density $\rho = 1190$
kg/m$^3$.
Figs.~\ref{branching}b and c show the fracture pattern at the
conclusion of the test. As may be seen, at a low prescribed
strain rate the crack tends to grow within its plane, and it
branches only when it senses the proximity of the free surface on
the right side of the specimen. Results from a similar calculation
at a higher nominal strain rate of 0.01/$\mu$s are shown in
Figs.~\ref{branchingbig}a-c. The mesh in this case contains
6,363 ten-node tetrahedra and 11,569 nodes, and the stable time
step is $\triangle t$ = 6 x 10$^-5$~$\mu$s.
Figs.~\ref{branchingbig}b and c show the crack patterns at the
conclusion of the test. Initially the crack remains within its
plane and accelerates steadily. As a certain crack speed is
attained, the crack begins to issue lateral branches. These
branches consume additional fracture energy, thereby limiting the
mean crack speed. As the crack approaches the free surface, the
extent of branching increases steadily.

As in the previous example, the ability of the method to deal with
complex geometrical and topological transitions simply and
effectively is noteworthy. In particular, cracks are allowed to
branch unimpeded, connect with free surfaces or with other cracks
to form fragments.

We conclude this section by emphasizing that the simulations
presented above, while representative of a broad class of
engineering materials and loading conditions, are not intended as
a validation of the cohesive model but as a demonstration of the
computational methodology. Detailed validation studies, including
extensive comparisons with experiment, based on test
configurations similar to those just described may be found
elsewhere \cite{pandolfi:1999, ruiz:2000, ruiz:2001, yuc:2001}.

\section{Summary and Conclusions}
\label{Summary}

In cohesive theories of fracture, material separation is governed
by a suitable cohesive law. In finite-element simulations based on
a tetrahedral triangulation of the domain of analysis, decohesion
and opening may conveniently be restricted to interior triangular
faces. The cohesive laws considered here are rigid up to the
attainment of the cohesive strength of the material.
Consequently, initially all the faces in the triangulation are
perfectly coherent, i.~e., conforming in the usual finite element
sense. Cohesive elements are inserted adaptively at interior faces
when the effective traction acting on those face reaches the
cohesive strength of the material. The insertion of cohesive
elements changes the geometry of the boundary and, frequently, the
topology of the model as well.

We have presented a simple set of data structures, and a
collection of methods for constructing and updating the
structures, designed to support the use of cohesive elements in
simulations of fracture and fragmentation. The data structures
and methods are straightforward to implement and enable the
efficient tracking of complex fracture and fragmentation
processes. The examples of application discussed here illustrate
the uncanny ability of the method to represent intricate
geometrical and topological transitions resulting from crack
branching, the nucleation of surfaces and interior cracks, crack
coalescence, the detachment of fragments, and others.

\section*{Acknowledgments}

The support of the Army Research Office through grant
DAA-H04-96-1-0056 is gratefully acknowledged. We are also grateful
for support provided by the DoE through Caltech's ASCI/ASAP
Center for the Simulation of the Dynamic Behavior of Solids. The
assistance provided by Dr.~Chengxiang Yu with the numerical
examples is gratefully acknowledged.

\bibliographystyle{unsrt}
{\small\bibliography{fragmen}}

\begin{thebibliography}{10}

\bibitem{field:1989}
J.E. Field, Q.~Sun, and D.~Townsend.
\newblock {Ballistic Impact of Ceramics}.
\newblock {\em Inst. Phys. Conf. Ser. No 102: Session 7, Paper presented at
  Int. Conf. Mech.Prop. Materials at High Rates of Strain, Oxford, 1989}, 1989.

\bibitem{kipp:1993}
M.~E. Kipp, D.~E. Grady, and J.~W. Swegle.
\newblock {Numerical and Experimental Studies of High-Velocity Impact
  Fragmentation}.
\newblock {\em {International Journal of Impact Engineering}}, 14:427--438,
  1993.

\bibitem{woodward:1994}
R.~L. Woodward, W.~A. Gooch, R.~G. O'Donnell, W.~J. Perciballi, B.~J. Baxter,
  and S.~D. Pattie.
\newblock {A Study of Fragmentation in the Ballistic Impact of Ceramics}.
\newblock {\em {International Journal of Impact Engineering}}, 15(5):605--618,
  1994.

\bibitem{piekutowski:1995}
A.J Piekutowski.
\newblock {Fragmentation of a Sphere Initiated by Hypervelocity Impact with a
  thin sheet}.
\newblock {\em {International Journal of Impact Engineering}}, 17:627--638,
  1995.

\bibitem{camacho:1996}
G.~T. Camacho and M.~Ortiz.
\newblock Computational modelling of impact damage in brittle materials.
\newblock {\em {International Journal of Solids and Structures}},
  33(20-22):2899--2938, 1996.

\bibitem{ortiz:1996}
M.~Ortiz.
\newblock {Computational Micromechanics}.
\newblock {\em {Computational Mechanics}}, 18:321--338, 1996.

\bibitem{pandolfi:1999}
A.~Pandolfi, P.~Krysl, and M.~Ortiz.
\newblock Finite element simulation of ring expansion and fragmentation: The
  capturing of length and time scales through cohesive models of fracture.
\newblock {\em {International Journal of Fracture}}, 95:1--18, 1999.

\bibitem{ruiz:2000}
G.~Ruiz, M.~Ortiz, and A.~Pandolfi.
\newblock Three dimensional finite-element simulation of the dynamic brazilian
  tests on concrete cylinders.
\newblock {\em {International Journal for Numerical Methods in Engineering}},
  48(7):963--994, 2000.

\bibitem{ruiz:2001}
G.~Ruiz, A.~Pandolfi, and M.~Ortiz.
\newblock Three-dimensional cohesive modeling of dynamic mixed-mode fracture.
\newblock {\em {International Journal for Numerical Methods in Engineering}},
  2001.
\newblock In press.

\bibitem{ortiz:1993}
M.~Ortiz and S.~Suresh.
\newblock {Statistical Properties of Residual Stresses and Intergranular
  Fracture in Ceramic Materials}.
\newblock {\em {Journal of Applied Mechanics}}, 60:77--84, 1993.

\bibitem{xux:1994}
X.~P. Xu and A.~Needleman.
\newblock {Numerical Simulations of Fast Crack Growth in Brittle Solids}.
\newblock {\em {Journal of the Mechanics and Physics of Solids}}, 42:1397,
  1994.

\bibitem{pandolfi:1998}
A.~Pandolfi and M.~Ortiz.
\newblock Solid modeling aspects of three-dimensional fragmentation.
\newblock {\em {Engineering with Computers}}, 14(4):287--308, 1998.

\bibitem{peraire:1987}
J.~Peraire, M.~Vahdati, K.~Morgan, and O.~C. Zienkiewicz.
\newblock {Adaptive Remeshing for Compressible Flow Computations}.
\newblock {\em {Journal of Computational Physics}}, 72:449--466, 1987.

\bibitem{peraire:1988}
J.~Peraire, J.~Peiro, L.~Formaggia, K.~Morgan, and O.C. Zienkiewicz.
\newblock {Finite Element Euler Computations in Three Dimensions}.
\newblock {\em {International Journal for Numerical Methods in Engineering}},
  26:2135--2159, 1988.

\bibitem{lohner:1988}
R.~L\"ohner and P.~Parikh.
\newblock {Generation of Three-Dimensional Unstructure Grids by the
  Advancing-Front Method}.
\newblock {\em {International Journal for Numerical Methods in Fluids}},
  8:1135--1149, 1988.

\bibitem{radovitzky:2000}
R.~Radovitzky and M.~Ortiz.
\newblock Tetrahedral mesh generation based on node insertion in crystal
  lattice arrangements and advancing-front-{Delaunay} triangulation.
\newblock {\em {Computer Methods in Applied Mechanics and Engineering}},
  187(3--4):543--569, 2000.

\bibitem{kane:1999}
C.~Kane, E.~A. Repetto, M.~Ortiz, and J.~E. Marsden.
\newblock Finite element analysis of nonsmooth contact.
\newblock {\em Computer Methods in Applied Mechanics and Engineering},
  180:1--26, 1999.

\bibitem{molinari:2001}
J.~F. Molinari and M.~Ortiz.
\newblock Three-dimensional adaptive meshing by subdivision and edge-collapse
  in finite-deformation dynamic-plasticity problems with application to
  adiabatic shear banding.
\newblock {\em {International Journal for Numerical Methods in Engineering}},
  2001.
\newblock submitted.

\bibitem{pandolfi:2001}
A.~Pandolfi, C.~Kane, M.~Ortiz, and J.~E. Marsden.
\newblock Time-discretized variational formulation of nonsmooth frictional
  contact.
\newblock {\em {International Journal for Numerical Methods in Engineering}},
  2001.
\newblock In press.

\bibitem{requicha:1980}
A.~A.~G. Requicha.
\newblock {Representations for Rigid Solids: Theory, Methods and Systems}.
\newblock {\em {Computing Surveys}}, 12:437--465, 1980.

\bibitem{mantyla:1988}
M.~Mantyla.
\newblock {\em {An Introduction to Solid Modeling}}.
\newblock Computer Science Press, Rockwille, Maryland, 1988.

\bibitem{hoffmann:1989}
C.~M. Hoffmann.
\newblock {\em {Geometric and Solid Modeling}}.
\newblock Morgan Kaufmann Publishers, San Mateo, California, 1989.

\bibitem{ortiz:1999}
M.~Ortiz and A.~Pandolfi.
\newblock A class of cohesive elements for the simulation of three-dimensional
  crack propagation.
\newblock {\em {International Journal for Numerical Methods in Engineering}},
  44:1267--1282, 1999.

\bibitem{deandres:1999}
A.~De-Andr\'es, J.~L. P\'erez, and M.~Ortiz.
\newblock Elastoplastic finite element analysis of three-dimensional fatigue
  crack growth in aluminum shafts subjected to axial loading.
\newblock {\em {International Journal of Solids and Structures}},
  36(15):2231--2258, 1999.

\bibitem{marsden:1983}
J.~E. Marsden and T.~Hughes.
\newblock {\em {Mathematical Foundations of Elasticity}}.
\newblock Prentice Hall, 1983.

\bibitem{belytschko:1983}
T.~Belytschko.
\newblock An overview of semidiscretization and time integration procedures.
\newblock In T.~Belytschko and T.~J.~R. Hughes, editors, {\em {Computational
  Methods for Transient Analysis}}, pages 1--65. North-Holland, 1983.

\bibitem{hughes:1983}
T.~J.~R. Hughes.
\newblock Analysis of transient algorithms with particular reference to
  stability behavior.
\newblock In T.~Belytschko and T.~J.~R. Hughes, editors, {\em {Computational
  Methods for Transient Analysis}}, pages 67--155. North-Holland, 1983.

\bibitem{yuc:2001}
C.~Yu, A.~Pandolfi, and M.~Ortiz.
\newblock 3d cohesive investigation on branching for brittle materials.
\newblock 2001.
\newblock In preparation.

\end{thebibliography}

\end{document}